\documentstyle[12pt,aaspp4]{article}
\begin{document}

\title{A POSSIBLE MASSIVE ASTEROID BELT AROUND $\zeta$ LEP}
\author{C. H. Chen and M. Jura}
\affil{Department of Physics and Astronomy, University of California,
       Los Angeles, CA 90095-1562; cchen@astro.ucla.edu; 
       jura@clotho.astro.ucla.edu}

\begin{abstract}
We have used the Keck I telescope to image at 11.7 $\mu$m and 17.9 $\mu$m
the dust emission around $\zeta$ Lep, a main sequence A-type star at 21.5 pc
from the Sun with an infrared excess. The excess is at most marginally 
resolved at 17.9 $\mu$m. The dust distance from the star is probably 
${\leq}$ 6 AU, although some dust may extend to 9 AU. The mass 
of observed dust is $\sim10^{22}$ g. Since the lifetime of dust particles is 
about $10^{4}$ yr because of the Poynting-Robertson effect, we robustly 
estimate at least $4\times10^{26}$ g must reside in parent bodies which may be 
asteroids if the system is in a steady state and has an age of $\sim$300 Myr. 
This mass is approximately 200 times that contained within the main asteroid 
belt in our solar system.
\end{abstract}


\section{INTRODUCTION}
Vega-like systems, first discovered by \emph{IRAS}, are main sequence
stars surrounded by dust. The operation of the Poynting-Robertson effect on
micron-sized grains, responsible for much of the observed infrared excess, 
requires that the lifetime of these particles be significantly shorter than 
the age of the stars \markcite{bp93}(Backman \& Paresce 1993). Thus, the dust 
grains must be replenished from a reservoir such as collisions between 
larger bodies or the sublimation of comets. Studies of Vega-like systems thus 
afford the opportunity to study the evolution of large solids such as planets, 
comets, and asteroids.

The identification of dust around main sequence A-type stars is usually made 
from the IRAS colors \markcite{mb98}\markcite{sil00}(e.g. Mannings \& Barlow 
1998; Silverstone 2000). Identifying objects with 12 $\mu$m excess is 
difficult because the photosphere usually dominates the total flux at this 
wavelength. The bulk of the dust associated with main sequence A-type stars 
typically has temperatures T$_{gr}$ $\sim$100 K, and semi-major axes $>$50 AU. 
For example, imaging of HR 4796A (A0V, distance from Earth 67.1 pc) in thermal 
infrared \markcite{krwb98}\markcite{jay98}(Koerner et al. 1998; Jayawardhana 
et al. 1998) and in near infrared scattered light \markcite{sch99}
(Schneider et al. 1999), revealed a dust ring with radius 70 AU, a distance 
scale similar to that of the Sun's Kuiper Belt. 

Dust grains located $\sim$5 AU from a main sequence A-type star have $T_{gr} 
\approx$ 300 K, producing strong 12 $\mu$m excesses. Searches for 12 $\mu$m 
excesses around main sequence stars have revealed that 300 K dust around 
A-K dwarfs is rare. In a survey of 548 A-K dwarfs, \markcite{ap91}Aumann \& 
Probst (1991) were able to identify IRAS 12 $\mu$m excesses 
only with $\beta$ Pictoris and $\zeta$ Lep (= HR 1998 = HD 38678). 
Similarly, \markcite{cot87}Cot\'{e} (1987) derived unusually high temperatures 
for the dust around $\zeta$ Lep compared to other Vega-like main sequence
stars from IRAS 12, 25 and 60 $\mu$m observations. Although coronagraphic 
images of $\beta$ Pic \markcite{kj95}(Kalas \& Jewitt 1995) have revealed a 
massive dust disk which extends to a radius of almost 1000 AU, no extended 
disk has been discovered around $\zeta$ Lep. We have carried out a high 
resolution mid-infrared imaging study to learn about the dust properties, and 
the origin and evolution of the dust around $\zeta$ Lep. 

$\zeta$ Lep is a main sequence A3Vn star with a \emph{Hipparcos} distance of 
21.5 pc and a fractional dust luminosity, 
L$_{IR}$/L$_{*}$ = 1.7$\times$10$^{-4}$ \markcite{ftk98}(Fajardo-Acosta, 
Telesco, \& Knacke 1998). It has been identified as a member of the Castor 
moving group, which also contains Fomalhaut and Vega, based upon its kinematic 
properties \markcite{byn98}(Barrado y Navascu\'{e}s 1998). 
\markcite{byn98}Barrado y Navascu\'{e}s (1998) estimates its age to be between 
100-400 Myr from the Lithium abundance of later-type stars in the Castor 
moving group and from the space motion of the group. Recently, 
\markcite{scb01}Song et al. (2001) estimated an age of 50-350 Myr  
using Stromgren $uvby\beta$ photometry corrected for the effects of rapid 
rotation, to determine stellar parameters (effective temperature $T_{eff}$,  
surface gravity $g$ and metallicity) which then are fit to stellar 
evolutionary tracks of \markcite{ssm92}Schaller et al. (1992). 
\markcite{ldl99}Lachaume et al. (1999) fit the position of $\zeta$ Lep on the 
H-R digram with theoretical isochrones of \markcite{ber94}Bertelli et al. 
(1994) and estimate a slightly older age of 200-500 Myr.

\section{OBSERVATIONS}
Our data were obtained on the night of 2001 February 5 (UT) at the Keck I
telescope using the Long Wavelength Spectrometer (LWS), which was built by a 
team led by B. Jones and is described on the Keck website. The LWS uses a 
128$\times$128 SiAs BIB array with a scale at the Keck telescope of 
0.08$\arcsec$ pixel$^{-1}$ and a total field of view of 
10.2$\arcsec\times$10.2$\arcsec$. We used the ``chop-nod'' mode of observing 
and two different filters: 11.2-12.2 $\micron$ and 16.9-18.9 $\micron$. We used
Capella ($\alpha$ Aur) for flux and point-spread function calibrations. 
The data were reduced at UCLA using standard LWS routines.

We flux calibrate our data using the results for Capella that 
$F_{\nu}$(11.5 $\mu$m) = 176.1 Jy and $F_{\nu}$(19.3 $\mu$m) = 67.5 Jy 
\markcite{gez87}(Gezari et al. 1987) extrapolated to our bands assuming that 
$F_{\nu} \propto \nu^{2}$ between 10 and 20 $\mu$m. For $\zeta$ Lep, we find 
F$_{\nu}$(11.7 $\mu$m) = 1.60$\pm$0.20 Jy and F$_{\nu}$(17.9 $\mu$m) = 
0.93$\pm$0.07 Jy in agreement with previous observations by 
\markcite{ftk98}Fajardo-Acosta et al. (1998). We conservatively estimate
the uncertainties associated with our measurements from the drift in the flux 
of Capella before and after our observations of $\zeta$ Lep. Since less time 
elapsed between our observations of the standard star at 17.9 $\mu$m, the 
percentage errors are smaller at this wavelength. 

Determining the infrared excess of $\zeta$ Lep requires subtracting a model for
the stellar photosphere from the observed fluxes. We model the spectral energy 
distribution (SED) of $\zeta$ Lep with a 1993 Kurucz stellar atmosphere, 
assuming solar metallicity ([Z/H]=0.0). We assume no interstellar extinction
because the star is only 21.5 pc from the Sun. We assume negligible 
circumstellar extinction since $L_{IR}/L_{*}$ $\sim$ $10^{-4}$ and we have no 
evidence in our data for an edge-on system. We find a best fit (minimum 
$\chi^{2}$) for the following parameters: stellar effective temperature 
T$_{eff}$ = 8500 K and surface gravity $\log g$ = 4.5 (see Figure 1). With 
this fit, we find photospheric fluxes of F$_{\nu}$(11.7 $\mu$m) = 
1.24$\pm$0.01 Jy and F$_{\nu}$(17.9 $\mu$m) = 0.53$\pm$0.01 Jy; thus, we find 
11.7 $\mu$m and 17.9 $\mu$m excesses of 0.36$\pm$0.2 Jy and 0.40$\pm$0.07 Jy 
respectively.

We convolve models for the dust distribution with the point spread function
(PSF) of Capella at 17.9 $\mu$m to estimate the maximum distance of the dust 
from the star. Observations of Capella made 40 minutes before, 20 minutes 
after and 100 minutes after our observations of $\zeta$ Lep varied somewhat. 
We use the PSF measured closest in time to our observations of $\zeta$ Lep to 
construct our analysis. We assume the the dust is confined to a face-on ring, 
with an average distance of 0.10$\arcsec$, 0.24$\arcsec$ or 0.40$\arcsec$ from 
the star and a width ranging between 0.20$\arcsec$ and 0.24$\arcsec$, and that 
55\% of the power is emitted by the point source and 45\% of the power is 
emitted by the ring, consistent with our model for the photosphere and the 
total flux measured from $\zeta$ Lep. We find a best fit for the model with 
the smallest dust disk (average radius 0.10\arcsec). The models including 
larger disks do not fit the surface brightness for the central pixel or the 
surface brightnesses for the outer pixels well. Thus, we estimate a maximum 
dust distance of 6 AU from the star. When models for the dust distribution are 
convolved with the PSF from earlier or later in the evening, the dust could 
extend as far as 9 AU from the star.

\section{DUST PROPERTIES}
\subsection{Minimum Grain Distance}
The minimum grain distance can be constrained from the temperature of the
grains assuming that they are black bodies. We estimate a grain temperature
of 320 K from the ratio of the 11.7 $\mu$m excess to the 17.9 $\mu$m excess.
Black bodies in radiative equilibrium with a stellar source are located
a distance
\begin{equation}
D = \frac{1}{2} (\frac{T_{*}}{T_{gr}})^{2} R_{*}
\end{equation}
from the central star \markcite{jur98}(Jura et al. 1998), where $T_{*}$ and 
$R_{*}$ are the effective temperature of the stellar photosphere and the 
stellar radius. We estimate the stellar 
luminosity from the bolometric magnitude using the \emph{Hipparcos} V-band 
magnitude ($m_{V}$ = 3.55 mag) and distance (21.5 pc) and a bolometric 
correction \markcite{flo96}(Flower 1996) corresponding to an effective 
temperature $T_{eff}$ = 8500 K. For $\zeta$ Lep, we find a stellar luminosity 
$L_{*}$ = 14 $L_{\sun}$. The stellar radius is therefore $R_{*}$ = 1.7 
$R_{\sun}$. From equation (1) and the stellar properties summarized 
in Table 1, we find a minimum grain distance of 2.8 AU. 

We can additionally infer the dust temperature and distance by fitting
a black body to the mid infrared fluxes reported for wavelengths longer than
10 $\mu$m. We find a best fit (minimum $\chi^2$) for the photosphere 
subtracted fluxes for $T_{dust}$ = 230 K and D = 5.4 AU, consistent with our 
observations.

\subsection{Circumstellar Dust Grain Size}
A lower limit to the size of dust grains orbiting a star can be found by
balancing the force due to radiation pressure with the force due to 
gravity. For small grains with radius $a$, the force due to radiation 
pressure overcomes gravity for:
\begin{equation}
a < 3 L_{*} Q_{pr}/(16 \pi G M_{*} c \rho_{s})
\end{equation}
\markcite{art88}(Artymowicz 1988) where $L_{*}$ and $M_{*}$ are the stellar 
luminosity and mass, $Q_{pr}$ is the radiation pressure coupling coefficient 
and $\rho_{s}$ is the density of an individual grain. Since radiation from 
an A-type star is dominated by optical and ultraviolet light, we expect that 
$2 \pi a/\lambda \gg 1$ and therefore the effective cross section of the 
grains can be approximated by their geometric cross section so 
$Q_{pr} \approx 1$. Based upon $T_{eff}$ and $L_{*}$, the inferred stellar 
mass is 1.9 $M_{\sun}$ \markcite{sdf00}(Siess, Dufour, \& Forestini 2000).
With $\rho_{s}$ = 2.5 g cm$^{-3}$, the minimum radius for grains orbiting 
$\zeta$ Lep is $a$ = 1.7 $\mu$m.

We can estimate the average size of the grains assuming a size distribution
for the dust grains. Analogous to the main asteroid belt in our 
solar system and as expected from equilibrium between production and 
destruction of objects through collisions \markcite{gn89}(Greenberg \& 
Nolan 1989), we assume
\begin{equation}
n(a) da = n_{o} a^{-p} da
\end{equation}
with p $\simeq$ 3.5 \markcite{bhs00}(Binzel, Hanner, \& Steel 2000). If we 
assume a minimum grain radius of 1.7 $\mu$m, we find an average grain radius 
$<a>$ = 2.8 $\mu$m, if we weight by the number of particles. 

\subsection{Mass of Circumstellar Dust Around $\zeta$ Lep}
We can estimate the minimum mass of dust around $\zeta$ Lep assuming that the 
particles have $<a>$ $\sim$ 2.8 $\mu$m; if the grains are larger, then our 
estimate is a lower bound. If we assume a thin shell of dust at 
distance, D, from the star and if the particles are spheres of radius, $a$, 
and if the cross section of the particles equals their geometric cross 
section, then the mass of dust is
\begin{equation}
M_{d} \geq \frac{16}{3} \pi \frac{L_{IR}}{L_{*}} \rho_{s} D^{2} <a> 
\end{equation}
\markcite{jur95}(Jura et al. 1995) 
where $L_{IR}$ is the luminosity of the dust. If D = 6 AU, 
$\rho_{s}$ = 2.5 g cm$^{-3}$, and $<a>$ = 2.8 $\mu$m, then $M_{d}$ = 
1.6$\times$10$^{22}$ g. 

\subsection{Lifetime of Circumstellar Grains}
One mechanism which may remove particles is Poynting-Robertson drag. 
The Poynting-Robertson lifetime of grains in a circular orbit, a distance $D$ 
from a star is
\begin{equation}
t_{PR} = \left( \frac{4 \pi <a> \rho_{s}}{3} \right) \frac{c^{2} D^{2}}{L_{*}}
\end{equation}
\markcite{bur79}(Burns et al. 1979). With the parameters given above and 
$L_{*}$ = 14 $L_{\sun}$, the Poynting-Robertson lifetime of the grains is 
$t_{PR}$ = 1.3$\times 10^{4}$ years. Since this timescale is significantly 
shorter than the stellar age ($t_{age}$), we hypothesize that the grains are 
replenished through collisions between larger bodies. By analogy with the 
solar system, we propose that the parent bodies are asteroids.

\subsection{Mass of the Parent Bodies}
We can estimate the total mass contained in parent bodies around $\zeta$ Lep
assuming a steady state. If $M_{PB}$ denotes the mass in parent bodies, then 
we may write
\begin{equation}
M_{PB} \geq \frac{M_{d}}{t_{PR}} t_{age} 
\end{equation}
With $M_{d}=1.6\times 10^{22}$ g, $t_{PR}=1.3\times10^{4}$ yr and 
$t_{age}=300\times10^{6}$ yr, we estimate a total mass of parent bodies  
$\sim4\times10^{26}$ g, approximately 200 times the mass of the main asteroid 
belt in our solar system (Binzel et al. 2000). From equations (4) and (5), we 
find
\begin{equation}
M_{PB} \geq \frac{4 L_{IR} t_{age}}{c^2}  
\end{equation}
Since L$_{IR}$ is well measured, our estimate for the lower limit on the mass 
of parent bodies, $M_{PB}$, is better constrained than our estimate for the 
mass of observed dust.

\section{DISCUSSION}
$\zeta$ Lep is distinctive because the dust around it is warm and close to the
star compared to the dust in other well known debris disk systems: $\beta$ 
Pictoris, HR 4796A, Vega, and Fomalhaut. We can quantify this difference by 
comparing the ratios of 60 $\mu$m excess to the 25 $\mu$m excess. For 
$\zeta$ Lep, this ratio is at least a factor of 2 smaller than for any of the 
other stars considered (see Table 2). 

The lack of dust at distances greater than 6 AU raises the possibility of the 
presence of a planet, sculping the disk and confining the dust. Such a planet
could increase the eccentricity of the orbits of the putative asteroids and 
thus drive them into mutual collisions to produce the observed dust.

\section{CONCLUSIONS}
We have obtained high resolution mid infrared images of $\zeta$ Lep at 11.7 
$\mu$m and 17.9 $\mu$m using the LWS on the Keck I telescope. 

1. $\zeta$ Lep  possess a strong mid infrared excess which is at most 
marginally resolved and thus surrounding dust probably lies within 6 AU
although some dust may extend as far as 9 AU. This result is consistent
with the temperature of the grains derived from the mid-infrared photometry.

2. Since the estimated Poynting-Robertson lifetime of grains with $<a>$ = 2.7 
$\mu$m is $1.3\times10^{4}$ yr, significantly shorter than the age of $\zeta$ 
Lep, the grains must be replenished from a reservoir such as collisions 
between larger bodies, perhaps asteroids. For $\zeta$ Lep, we robustly 
estimate that the minimum mass of parent bodies is $\sim4\times10^{26}$ g 
assuming that the system is in a steady state. This mass is approximately 200 
times the mass of the Sun's main asteroid belt.

This work has been supported by funding from NASA. We thank M. Sykes, A.
Weinberger, and B. Zuckerman for their comments.

\newpage
\begin{table}
\begin{center}
\footnotesize
Table 1. $\zeta$ Lep Properties 
\\
\begin{tabular}{lll} \tableline \tableline
    Quantity & Adopted Value & Reference \\ \tableline
    Spectral Type & A3Vn & 1 \\
    Distance & 21.5 pc & 2 \\
    Effective Temperature (T$_{eff}$) & 8500 K & \\
    Surface Gravity ($\log g$) & 4.5 & \\
    Stellar Radius (R$_{*}$) & 1.7 R$_{\sun}$ & \\
    Stellar Luminosity (L$_{*}$) & 14 L$_{\sun}$ & \\
    Rotational Velocity ($v\sin i$) & 202 km/sec & 1 \\
    Fractional Dust Luminosity & 1.7$\times$10$^{-4}$ & \\
        \ \ \ \ \ ($L_{IR}/L_{*}$) & & \\
    Age & 50-500 Myr & 3, 4, 5 \\ \tableline
\end{tabular}
\tablerefs{(1) \markcite{hw91}Hoffleit \& Warren 1991;
           (2) \emph{Hipparcos};
           (3) \markcite{byn98}Barrado y Navacu\'{e}s 1998;
           (4) \markcite{ldl99}Lachaume et al. 1999;
           (5) \markcite{scb01}Song et al. 2001}
\end{center}
\end{table}
\vspace{1mm}
\normalsize

\newpage
\begin{table}
\begin{center}
\footnotesize
Table 2. Disk Properties 
\\
\begin{tabular}{lccc} \tableline \tableline
    Star & Photosphere+Excess\tablenotemark{a} & Excess\tablenotemark{b} &
        T$_{color}$\tablenotemark{c} \\
    & $\frac{F_{\nu}(60 \mu m)}{F_{\nu}(25 \mu m)}$ &
        $\frac{F_{\nu}(60 \mu m)}{F_{\nu}(25 \mu m)}$ & (K) \\ \tableline
    $\zeta$ Lep & 0.32 & 0.30 & 370 \tablenotemark{d} \\
    $\beta$ Pic & 2.2 & 2.3 & 100 \\
    HR 4796A & 2.2 & 1.8 & 110 \\
    Vega & 0.85 & 4.4 & 80 \\
    Fomalhaut & 2.0 & 11 & 70 \\ \tableline
\end{tabular}
\tablenotetext{a}{IRAS data without color corrections} 
\tablenotetext{b}{IRAS data with color corrections and photospheric 
    subtraction}
\tablenotetext{c}{derived from F$_{\nu}$(60 $\mu$m)/F$_{\nu}$(25 $\mu$m) for
            the excess radiation given in the previous column}
\tablenotetext{d}{a black body plus stellar photosphere fit to all of the 
    fluxes reported for wavelengths greater than 10 $\mu$m yields a lower 
    temperature, $T_{gr}$ = 230 K}
\end{center}
\end{table}
\vspace{1mm}
\normalsize

\newpage

\begin{figure}[ht]
\figurenum{1}
\epsscale{0.80}
\plotone{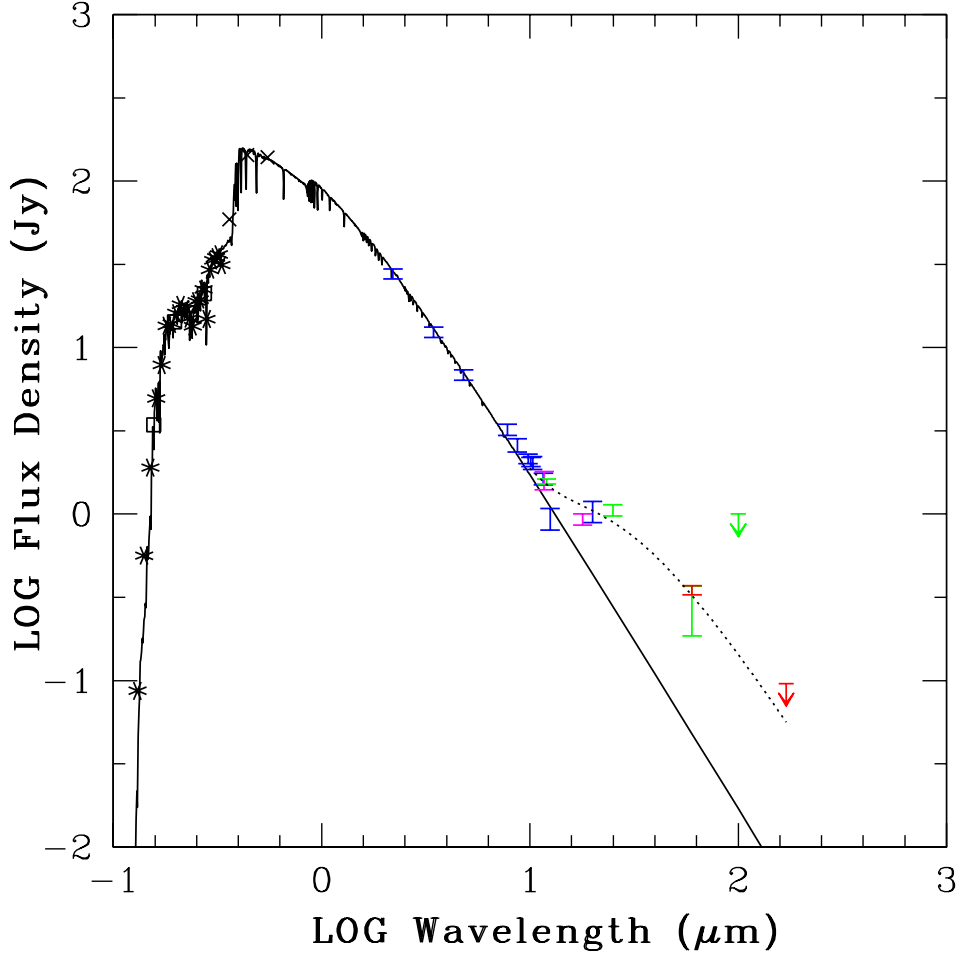}
\figcaption{Spectral energy distribution (SED) for $\zeta$ Lep with 
ultra-violet fluxes from TD 1 (Thompson et al. 1978) plotted as squares and 
IUE plotted as asterisks and UBV photometry (Cousins 1984) plotted as crosses. 
Previous JHK and mid-infrared photometry from the IRTF (Fajardo-Acosta et al. 
1998) are shown with dark blue error bars, from the IRAF Faint Source Catalog 
(Moshir et al. 1989) with green error bars and from ISO (Habing et al. 2001) 
with red error bars. Our 11.7 $\mu$m and 17.9 $\mu$m photometry, as reported 
here, is shown with magenta error bars. Overlaid is a 1993 Kurucz model for a 
stellar atmosphere with $T_{eff}$ = 8500 K, surface gravity $\log g$ = 4.5 and 
solar metallicity ([Z/H]=0.0) plotted to minimize $\chi^2$. The dotted black
line is a best fit to the mid infrared fluxes at wavelengths longer than 
10 $\mu$m assuming a single temperature black body (T = 230 K) plus the 
stellar photosphere.}
\label{1}
\end{figure}

\begin{figure}[ht]
\figurenum{2a}
\plotone{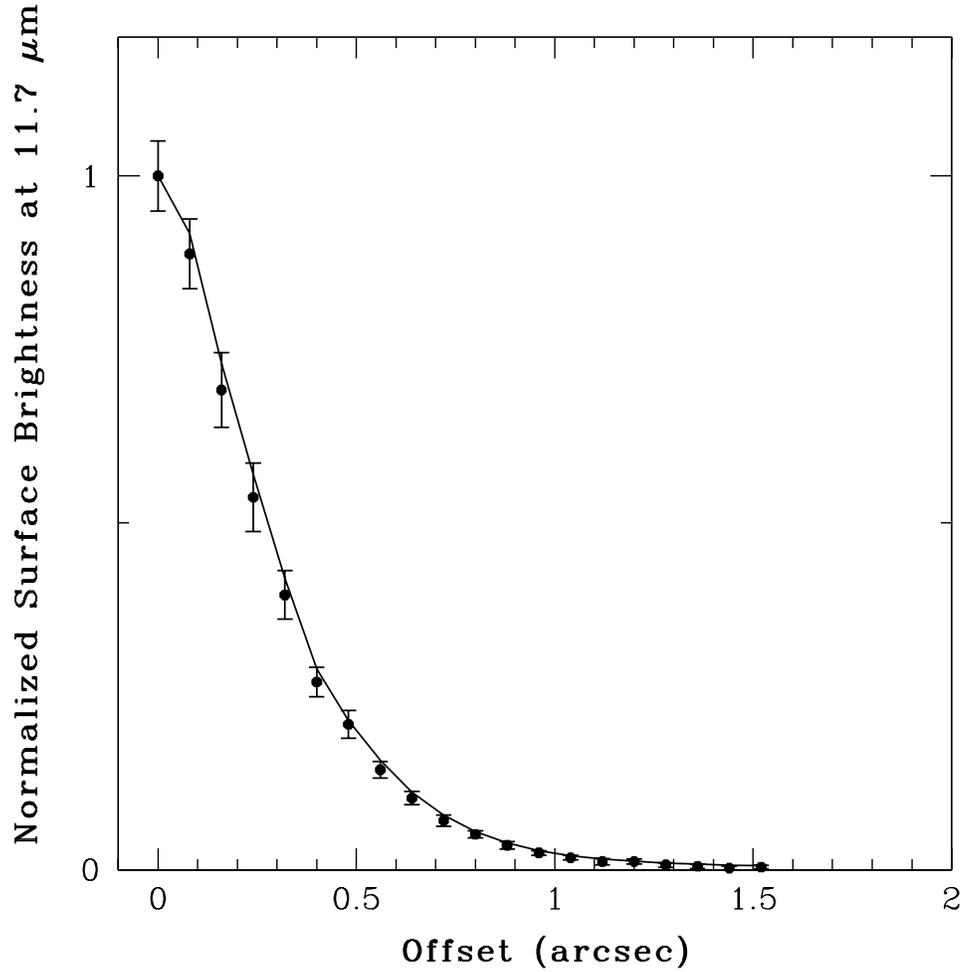}
\figcaption{Radial profile for $\zeta$ Lep 
    ($\alpha$(J2000.0)=$05^{h}46^{m}57^{s}$,
    $\delta$(2000.0)=$-19\arcdeg 49\arcmin 19\arcsec$) at 11.7 $\mu$m. 
    The solid black line is the point spread function for the standard star 
    Capella ($\alpha$(J2000.0)=$05^{h}16^{m}41^{s}$,
    $\delta$(2000.0)=$+45\arcdeg 59\arcmin 53\arcsec$) 
    normalized to the observations of $\zeta$ Lep, while the error bars show 
    the azimuthally averaged data for $\zeta$ Lep.}
\end{figure}

\begin{figure}[ht]
\figurenum{2b}
\plotone{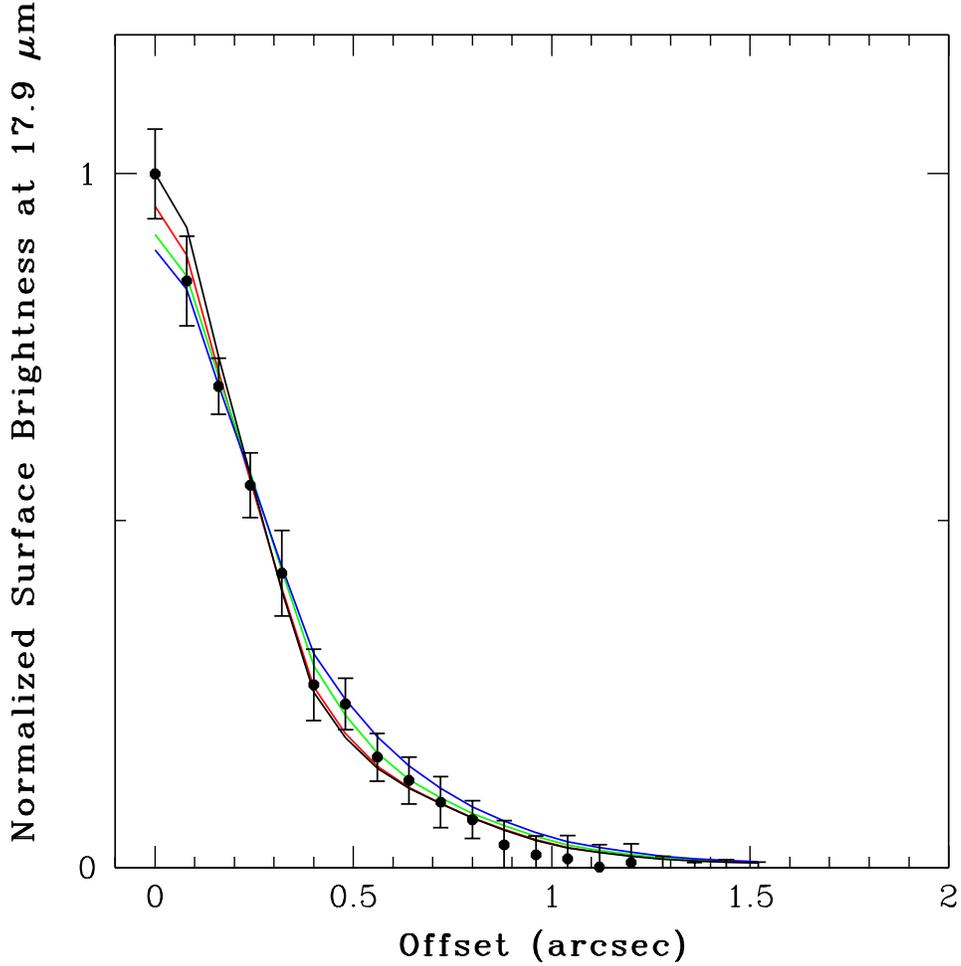}
\figcaption{ Radial profile for $\zeta$ Lep at 17.9 $\mu$m. The solid black line 
    is the point spread function for the standard star Capella, normalized to 
    the observations of $\zeta$ Lep, while the error bars show the azimuthally 
    averaged data for $\zeta$ Lep. The colored lines are best fit models 
    (minimum $\chi^2$) for the spatial extent of the dust convolved with the 
    point spread function of Capella. The red, green and blue models assume a 
    central point source and a ring with average angular radius 0.10$\arcsec$, 
    0.24$\arcsec$ and 0.40$\arcsec$ respectively. The $\chi^2$ values for the 
    different curves here are 4.5, 5.2, and 9.3. In each model, 55\% of the 
    power is emitted by the point source and 45\% of the power is emitted by 
    the ring, consistent with our model for the photosphere and the total flux 
    emitted from $\zeta$ Lep.}
\end{figure}

\end{document}